\documentclass[
 reprint, 10pt,
 amsmath,amssymb,
 aps, superscriptaddress
]{revtex4-1}

\usepackage{graphicx}
\usepackage{dcolumn}
\usepackage{hyperref}
\usepackage[dvipsnames]{xcolor}
\usepackage{bm}
\usepackage{bbold}
\usepackage{bbm}

\definecolor{marie}{RGB}{0,128,128}
\definecolor{todo}{RGB}{102,51,153}
\definecolor{emph}{RGB}{0,150,150}

\begin{document}

\title{Theory of strong coupling between molecules and surface plasmons on a grating}

\author{Marie S Rider}
\email{M.S.Rider@exeter.ac.uk}
\affiliation{Department of Physics and Astronomy, Stocker Road, University of Exeter, Devon EX4 4QL, United Kingdom}
\author{Rakesh Arul}
\affiliation{NanoPhotonics Centre, Department of Physics, University of Cambridge, Cambridge CB3 0HE, United Kingdom}
\author{Jeremy J Baumberg}
\affiliation{NanoPhotonics Centre, Department of Physics, University of Cambridge, Cambridge CB3 0HE, United Kingdom}
\author{William L Barnes}
\email{W.L.Barnes@exeter.ac.uk}
\affiliation{Department of Physics and Astronomy, Stocker Road, University of Exeter, Devon EX4 4QL, United Kingdom}

\date{\today}
\begin{abstract}
The strong coupling of molecules with surface plasmons results in hybrid states which are part molecule, part surface-bound light. Since molecular resonances may acquire the spatial coherence of plasmons, which have mm-scale propagation lengths, strong-coupling with molecular resonances potentially enables long-range molecular energy transfer. Gratings are often used to couple incident light to surface plasmons, by scattering the otherwise non-radiative surface plasmon inside the light-line. We calculate the dispersion relation for surface plasmons strongly coupled to molecular resonances when grating scattering is involved. By treating the molecules as independent oscillators rather than the more typically-considered single collective dipole, we find the full multi-band dispersion relation. This approach offers a natural way to include the dark states in the dispersion. We demonstrate that for a molecular resonance tuned near the crossing point of forward and backward grating-scattered plasmon modes, the interaction between  plasmons and molecules gives a five-band dispersion relation, including a bright state not captured in calculations using a single collective dipole. We also show that the role of the grating in breaking the translational invariance of the system appears in the position-dependent coupling between the molecules and the surface plasmon. The presence of the grating is thus not only important for the experimental observation of molecule-surface-plasmon coupling, but also provides an additional design parameter that tunes the system. 
\end{abstract}
\maketitle

\section{Introduction}
\label{sec:introduction}
The strong coupling of molecules and light is a topic of rapidly increasing interest, due to its potential application in multiple fields including quantum computing, photocatalysis and photovoltaics. Multiple recent reviews and perspectives exist on this emerging topic, see for example~\cite{torma2014strong,herrera2020molecular,yuen2022polariton}. An active area of research is the control of molecules coupled to surface plasmons, where there is an opportunity to control inter-molecular energy transfer over length scales commensurate with the propagation length of the surface plasmons; it is hoped that strong coupling may extend the range of molecular energy transfer from nanometres to millimetres. Strong coupling between molecular resonances and surface plasmons is a vacuum effect, so it is not necessary to inject light into the system. However, surface plasmons are usually non-radiative because they have more momentum along the surface than a photon of the same frequency, and thus lie outside the light-line. Momentum matching is thus required to observe the effect of molecules on the surface plasmons arising from strong coupling. The original but still convenient and powerful approach is to introduce a periodic modulation of the interface at which the surface plasmons propagate~\cite{wood1902xlii,Ritchie_PRL_1968_21_1530}, an approach which is still very topical~\cite{dintinger2005strong,rivas2008surface,jiang2011surface,niu2015image,wood2017single}. 
\\

Here we make use of a microscopic quantum electrodynamics (QED) model to analyse in detail the strong coupling interaction between molecular resonances and surface plasmons in the presence of a periodically modulated (grating) surface. Throughout we use the full $X+N$ model, where $X$ is the number of photonic modes and $N$ is the number of molecules. We first provide a reference point by employing our approach to a more familiar strong coupling system - molecules coupled to the cavity mode of a planar optical microcavity. We then discuss molecules coupled to a surface plasmon, then introduce the grating. We find the $(X+N)\times(X+N)$ Hamiltonian and corresponding dispersion relation of each system, and demonstrate that for molecules coupling to SPs on a grating, the spatial distribution of the molecules relative to the grating provides an additional control feature with which to tune the system. We contrast our results to those found by modelling the molecules via a single collective dipole mode (i.e. a $X+1$ model) and rigorous coupled wave analysis (RCWA), and demonstrate that the $X+N$ model is necessary to capture the full character of the system. 
%

\section{Strong coupling of molecules to a cavity mode}
\label{sec:SC_molecules_cavity}
We set the scene with the familiar system of molecules in a single-mode planar optical cavity. For simplicity we assume only a single molecular resonance (it could be excitonic, vibronic etc.) that lies within the frequency range of interest. The excitation of molecule $i$ with transition energy $\hbar \omega_{\mathrm{mol}}$ can be described in the low-excitation regime (see Appendix~\ref{app:low_ex}) by a Fock state $|n_i\rangle$, where $n_i=0,1$. The Fock state is operated on with the bosonic creation and annihilation operators $b^\dagger_i,b_i$. Each molecular transition dipole is given by,
\begin{align}
    \bm{\mu}_i &= \mathbf{M}_i b^\dagger_i+ \mathbf{M}^*_i b_i, 
\end{align}
where $\mathbf{M}_i=e_0 \langle e| \mathbf{d}_i|g\rangle$ is the $i$-th transition dipole matrix element, $\mathbf{d}_i$ is the displacement vector of the $i$-th dipole, and $b^\dagger_i,b_i$ act on the $i$-th molecule. 
\\

\begin{figure}
    \centering
    \includegraphics[width=\columnwidth]{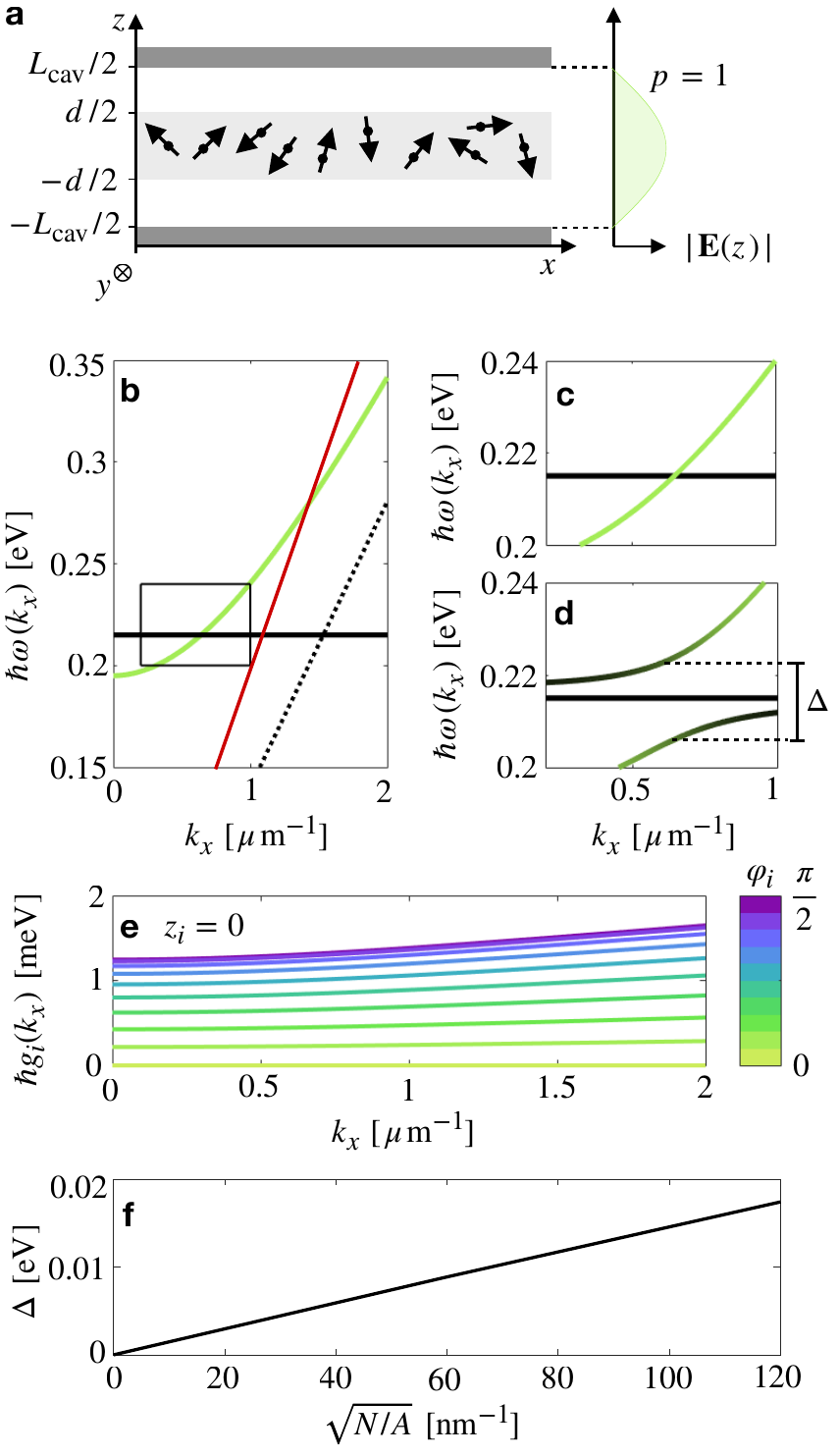}
    \caption{\textbf{Strong coupling of molecules to a planar cavity mode:} \textbf{(a)} Schematic showing a slab (of thickness $d$) of transition dipoles in a cavity of thickness $L_\mathrm{cav}$ and \textbf{(b)} the dispersion relation of $N$ C=O units, with transition resonance $\hbar\omega_{\mathrm{mol}}=0.215$ eV, and a cavity ($L_\mathrm{cav}=2.26~\mu$m,) mode, with no molecule/cavity coupling. Light-line in air given by red line, light-line in PMMA given by dotted black line. \textbf{(c)} Zoom-in of overlap between modes without and \textbf{(d)} with strong coupling. PMMA density $N/A = 1.58\times 10^{4}$ nm$^{-2}$, slab thickness $d=L_\mathrm{cav}=2.26$ $\mu$m, dipoles aligned with electric field such that all $\vartheta_i=\varphi_i=\pi/2$. $\Delta$ is the magnitude of the gap at the resonant frequency. The dark states can be seen as a dispersionless band of frequency $\omega_\mathrm{mol}$. \textbf{(e)} Magnitude of the coupling constant for a  molecule at position $z_i=0$ and varying dipole alignment from along the x axis (such that $\varphi = 0$, $\vartheta=\pi/2$) to along the y axis (such that  $\varphi = \pi/2$, $\vartheta=\pi/2$). \textbf{(f)} Varying the interaction strength by varying PMMA density $N/A$, we see that the size of the avoided crossing gap is proportional to $\sqrt{N/A}$.}
    \label{fig:SC_molecules_cavity}
\end{figure}

The Hamiltonian of $N$ uncoupled molecules (i.e. assuming no interactions between molecules) is given by, 
\begin{align}
    \mathcal{H}_{\mathrm{mol}} &= \hbar \omega_{\mathrm{mol}}\sum_{i=1}^N b^\dagger_i b_i.
\end{align}
We consider the molecules to reside within a slab of thickness $d$ and area $A$, of dielectric constant $\varepsilon_2$, with the $i$-th molecule positioned at $(x_i,y_i,z_i)$. The molecules are randomly spaced with density $N/V$, where $V=Ad$. We place the slab containing the molecules in a cavity comprised of two mirrors (as illustrated in FIG.~\ref{fig:SC_molecules_cavity}a), perpendicular to the $z$-axis. The cavity modes are given by, 
\begin{align}
\omega_{\mathrm{cav},p}(k_x,k_y) = \frac{c}{n}\sqrt{ \left(\frac{p \pi}{L_\mathrm{cav}}\right)^2 + k_x^2+k_y^2}. 
\end{align}
We assume the $\omega_{\mathrm{cav},p}$ is real in this work for simplicity, but could be made complex to include absorption in the mirrors and leakage from the cavity. Assuming the molecular material fully fills the cavity such that $d=L_\mathrm{cav}$, $n=\sqrt{\epsilon_2}$. We assume the cavity thickness, $L_\mathrm{cav}$, is such that only the lowest-order mode can propagate and is launched in the $x$-direction, such that, 
\begin{align}
    \omega_{\mathrm{cav}}(k_x)=\omega_{\mathrm{cav},1}(k_x) = \frac{c}{n}\sqrt{ \left(\frac{ \pi}{L_\mathrm{cav}}\right)^2 + k_x^2}.
\end{align}
Considering a TE-polarised field, the cavity field is then given by,
\begin{align}
    \mathbf{E}&= i \left(\frac{\hbar\omega_{\mathrm{cav}}(k_x)}{2 \epsilon_2\epsilon_0 V}\right)^{\frac{1}{2}} \mathrm{cos}\left(\frac{\pi z}{L_\mathrm{cav}}\right)\mathbf{e}_y \left[a e^{i k_x x}+ a^\dagger e^{-i k_x x}  \right],
    \label{eq:electric_field_cav}
\end{align}
where $a^\dagger,a$ are the bosonic creation and annihilation operators of the cavity field. $V=A L_\mathrm{cav}$ is the volume of the cavity of thickness $L_\mathrm{cav}$ and $A$ is the same area as that of the molecular slab. The form of the electric field in Equation~\ref{eq:electric_field_cav} is comprised of factors relating to: the field strength, the spatial distribution, and the propagation behaviour respectively. The Hamiltonian for the cavity mode is,
\begin{align}
    \mathcal{H}_{\mathrm{cav}} & = \hbar \omega_{\mathrm{cav}}(k_x) a^\dagger a.
\end{align}
The molecular transition dipoles interact with the cavity mode via the light-matter interaction Hamiltonian, which - assuming the resonant electric dipole approximation - is given by,
\begin{align}
    \mathcal{H}_{\mathrm{int}} &= - \sum_{i=1}^N\bm{\mu}_i\cdot \mathbf{E} \\
     &=   \hbar \sum_{i=1}^N   \left[ g_{i}(k_x) b^\dagger_i a  + g_{i}^*(k_x) b_i a^\dagger \right] ,
\end{align}
where we have neglected non-energy-conserving terms (involving $b^\dagger_i a^\dagger$ and $b_i a$, note though that it would be necessary to keep them if we want to explain higher-order processes), and the interaction strength $g_{i}(k_x)$ is given by, 
\begin{align}
    g_{i}(k_x) &= i \left(\frac{\omega_{\mathrm{cav}}(k_x)}{2\hbar \epsilon_2\epsilon_0 V}\right)^{\frac{1}{2}}  \mathrm{cos}\left(\frac{\pi z_i}{L_\mathrm{cav}}\right)e^{i k_x x_i} \mathbf{M}_i\cdot \mathbf{e}_{y}.
\end{align}
Here we have assumed that the wavelength of light is much larger than the spatial extent of an individual molecule, and so for the $i$-th term we can rewrite $e^{ik_x x} = e^{ik_x (x-x_i) }e^{ik_x x_i }$, and for $k_x (x-x_i)\ll 1$, $ e^{ik_x (x-x_i)}=1+ i k_x (x-x_i)+... \approx 1$, and so $e^{ik_x x} \approx  e^{ik_x x_i}$. Each molecule only sees the field at its position. By similar reasoning, $\mathrm{cos}(\pi z/L_\mathrm{cav})\approx \mathrm{cos}(\pi z_i/L_\mathrm{cav})$. The interaction strength varies with the $z$ position, as has been demonstrated experimentally~\cite{ahn2018vibrational}. This in turn affects the final dispersion relation of the system, while the $x$ position of the molecules does not. 
\\

We may write the transition dipole matrix element as 
\begin{align}
    \mathbf{M}_i = |\mathbf{M}|\begin{pmatrix}\mathrm{cos}\varphi_i \mathrm{sin}\vartheta_i\\ \mathrm{sin}\varphi_i \mathrm{sin}\vartheta_i \\\mathrm{cos}\vartheta_i\end{pmatrix},
    \label{eq:mol_orientation}
\end{align}
where we have assumed that all molecular dipoles have the same magnitude but are independent in orientation. The interaction strength is then, 
\begin{align}
    g_{i}(k_x) &= i \left(\frac{\omega_{\mathrm{cav}}(k_x)}{2\hbar \epsilon_2\epsilon_0 V}\right)^{\frac{1}{2}}\mathrm{cos}\left(\frac{\pi z_i}{L_\mathrm{cav}}\right) e^{i k_x x_i}|\mathbf{M}|  \mathrm{sin}\varphi_i \mathrm{sin}\vartheta_i.
\end{align}
%

The total Hamiltonian of the system is given by
\begin{align}
\begin{split}
    \mathcal{H} &= \mathcal{H}_{\mathrm{cav}}+\mathcal{H}_{\mathrm{mol}}+\mathcal{H}_{\mathrm{int}} 
\end{split}\\
\begin{split}
    &= \hbar \omega_{\mathrm{cav}}(k_x) a^\dagger a + \hbar \omega_{\mathrm{mol}}\sum_{i=1}^N b^\dagger_i b_i \\
    &\quad+ \hbar \sum_{i=1}^N   \left[ g_{i}(k_x) b^\dagger_i a  + g_{i}^*(k_x) b_i a^\dagger \right].
\end{split}
\end{align}
It is possible to define a collective coupling strength for $N$ molecules, $g_N = \sqrt{\sum_i{|g_i|^2}}\propto \sqrt{N/A}$~\cite{gonzalez2013theory}, resulting in a single term in the interaction Hamiltonian (+h.c.), and a $2\times2$ matrix for the total Hamiltonian. This is particularly useful for large $N$, which is a sensible assumption in most experimental situations. Representing the molecules as a single effective dipole is an elegant way to study the two polaritons expected in the system, however it misses out the remaining molecular dark states. In a cavity system this is not of great importance as the dark states are typically not seen owing to the high impedance mismatch that prevents much light entering the cavity around the resonance frequency~\cite{houdre1994measurement,skolnick1998strong}. This high impedance is in turn due to the strong absorption associated with the molecular resonance. However in plasmonic systems we know from experiment~\cite{menghrajani2019hybridization,menghrajani2019vibrational} that these modes are often seen so that here we retain a full treatment of the $N$ individual molecular modes so as to accurately capture the multi-band features of the resulting system. However, to avoid unduly long computation times associated with diagonalising our $X+N$ Hamiltonian we choose a value of $A$ such that we can obtain molecular densities commensurate with typical experimental results. While $N$ should be small enough to give a short calculation time, it must also be large enough to yield the expected bulk response, which is particularly important if the molecules are taken to have random spatial distribution. $N$ should be larger than the number of distinct bands in the system, and values of $N\approx 100$ qualitatively gives the bulk result, given randomly distributed molecules. Larger values of $N$ can be taken to fully converge to the bulk system. 
We can write the Hamiltonian in matrix form as,
\begin{align}
    \mathcal{H}&= \hbar\begin{pmatrix}a^\dagger,b^\dagger_1,b^\dagger_2,...b^\dagger_N\end{pmatrix}
    \mathbf{H}
    \begin{pmatrix}a \\b_1 \\b_2 \\ \vdots \\ b_N \end{pmatrix},
\end{align}
where $\mathbf{H}$ is a $(1+N) \times (1+N)$ matrix,
\begin{align}
    \mathbf{H} = \begin{pmatrix}
    \omega_\mathrm{cav}(k_x) & g_1(k_x) & g_2(k_x) & \dots & g_N(k_x) \\
    g_1^*(k_x) & \omega_{\mathrm{mol}} & 0 & \dots & 0 \\
     g_2^*(k_x)  & 0 & \omega_{\mathrm{mol}}  & \dots & 0 \\
    \vdots & \vdots & \vdots & \ddots & \vdots \\
    g_N^*(k_x)  & 0 & 0 & \dots & \omega_{\mathrm{mol}} 
    \end{pmatrix}.
\end{align}

We now study a particular molecule-cavity system in order to discuss the dispersion relation. We choose to study the stretching vibration of the C=O bond of PMMA (polymethyl methacrylate, commonly known as acrylic). The transition matrix dipole moment is oriented along the bond, and the resonance of this excitation occurs at $\hbar\omega_\mathrm{mol} = 0.215$~eV. The bulk dielectric constant of PMMA is taken to be $\epsilon_2=1.99$. We choose a cavity thickness of $L_\mathrm{cav}=2.26~\mu$m such that the cavity mode and the molecular resonance cross as shown in FIG.~\ref{fig:SC_molecules_cavity}b, with the cavity dispersion relation shown in green, and the molecular resonance in black. The dotted black line is the light-line in PMMA, and the solid red line is the light-line in air. The approximate value of the C=0 dipole moment is given by $|\mathbf{M}|=8.0\times 10^{-31}$ Cm, found by fitting to experimental data for PMMA in a cavity~\cite{menghrajani2019hybridization}. This extracted value of the dipole moment will then be used in the next sections, in which PMMA is placed in other photonic environments - namely on metallic slabs and gratings. As the density of PMMA is $\rho=1.18$g$/$cm$^{3}$, $N/V = 7/$nm$^3$ and for a slab which fills the cavity such that $d=2.26$~$\mu$m, we thus set $N/A = 1.58\times10^4/$nm$^2$.
\\

\begin{figure}
    \centering
    \includegraphics[width=\columnwidth]{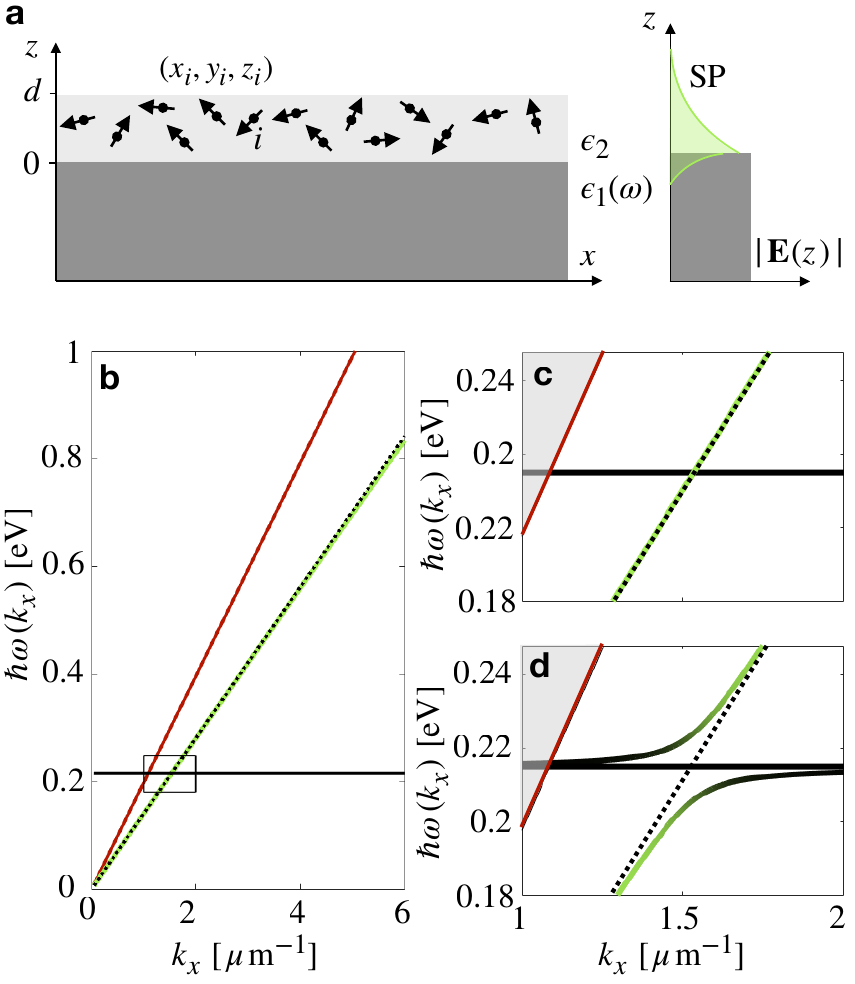}
    \caption{\textbf{Strong coupling of molecules to a surface plasmon:} \textbf{(a)} Schematic showing transition dipoles in a slab of thickness $d$ on a metal plate which we consider to have infinite thickness. An SP can propagate at the metal/dielectric interface. \textbf{(b)} The dispersion relations of an SP at a gold/PMMA interface and $N$ uncoupled C=O units. The red line gives the light-line in air, and the dotted line gives the light-line in PMMA. \textbf{(c)} Zoom-in of the region where the dispersion relations overlap, and \textbf{(d)} the dispersion relation with strong coupling. $N/A=70$~nm$^{-2}$, d = $2~\mu$m.}
    \label{fig:SC_molecules_SPP}
\end{figure}
Diagonalising $\mathbf{H}$, we find the dispersion relation of the coupled system. FIG.~\ref{fig:SC_molecules_cavity}c shows the relevant section of the uncoupled dispersion relation. FIG.~\ref{fig:SC_molecules_cavity}d gives the dispersion relation for the maximal interaction strength where all transition dipoles are oriented in the $y$ direction, such that $\varphi_i=\vartheta_i=\pi/2$. We see an avoided crossing at the resonance frequency, with the gap between the upper and lower hybrid modes $\Delta = 2 \hbar \sqrt{\sum_i |g_i(k_{x,\Delta})|^2}$, where $g_i(k_{x,\Delta})$ is the interaction strength of the $i$-th molecule at the crossing point of the two dispersion relations. The polariton modes demonstrate a mixing of photonic and molecular character, as shown by their colour projection (where green gives the projection of the hybrid state onto the unperturbed cavity mode and black gives the projection onto the molecular mode). In total there are three distinct modes - two polariton modes and $N-1$ degenerate molecular modes; modes that have not taken on any photonic properties. The interaction strength of each molecule can be tuned through the orientation of its transition dipole, with the maximal value for orientation along the $y$ axis, and complete decoupling when oriented along the $x$-axis, i.e. when $\varphi_i=\vartheta_i=0$, as illustrated in FIG.~\ref{fig:SC_molecules_cavity}e for a molecule with $z=0$. FIG~\ref{fig:SC_molecules_cavity}e also highlights the $k_x$ dependence of the interaction strength. $k_x$ enters the interaction strength via both the field strength (as mode frequency is $k_x$-dependent), and also through a phase term. In practice this dependence is often neglected, particularly in systems with losses, as it is the relative strengths of terms in the Hamiltonian which will dominate the characteristics of the system. In FIG.~\ref{fig:SC_molecules_cavity}f we plot $\Delta$ against $\sqrt{N/A}$ for fixed $d=2.26~\mu$m, from which we see the dependence of the gap on the density of molecules.
\\

Having established our approach in the context of strong coupling for molecules within a planar cavity, in Section~\ref{sec:SC_molecules_SPP} we look at molecules coupling to surface plasmons at a dielectric/metal interface. With their near-surface confinement, and their long propagation lengths, plasmons present additional features to those found in cavity modes which could be highly useful when integrated into polaritonic systems with molecules. Once we have established the situation for the planar system we will then add the final ingredient, the grating, in Section~\ref{sec:SC_molecules_SPP_grating}.

\section{Strong coupling of molecules to a surface plasmon}
\label{sec:SC_molecules_SPP}

Surface plasmons (SPs)~\footnote{strictly they are surface plasmon-polaritons~\cite{Burstein_Polaritons_Intro_1974}} are highly confined electromagnetic modes which propagate at dielectric/metal interfaces. It is the negative permittivity of the metal (the conduction electrons in the metal behave as a plasma, and have an associated plasma frequency, $\omega_0$) that enables this behaviour. For a SP launched in the $x$ direction on a metallic slab (see FIG.~\ref{fig:SC_molecules_SPP}a), the SP dispersion is given by~\cite{kloos1968dispersion,kretschmann1968radiative},
\begin{align}
    k_x(\omega_{\mathrm{SP}}) = \frac{\omega_{\mathrm{SP}}}{c}\sqrt{\frac{\epsilon_1 (\omega_{\mathrm{SP}})\epsilon_2 }{\epsilon_1 (\omega_{\mathrm{SP}})+\epsilon_2 }},
\end{align}
where $\epsilon_2$ is the background dielectric constant and we model the dielectric function of the metallic slab with a simple Drude-Lorentz model,
\begin{align}
    \epsilon_1 (\omega) = 1-\frac{\omega_0^2}{\omega^2+ i \gamma \omega},
\end{align}
where for gold, the plasma frequency $\omega_0=1.29\times10^{16}$~rad~s$^{-1}$ and damping is given by $\gamma = 7.3\times10^{13}$~rad~s$^{-1}$. We assume a slab of thickness $d$ contains the molecules with density $N/V$, where $V=A L_z$ and $L_z$ is the penetration depth of the plasmonic electric field into the dielectric, which is of order the wavelength of light in the dielectric, and for PMMA we take to be $L_z\sim 2.9~\mu$m.  FIG.~\ref{fig:SC_molecules_SPP}b gives the dispersion relation of the SP (green solid line) and the molecular resonance of PMMA (black solid line), as used in the previous section. It is important to note that the SP mode, and thus the crossover of the SP and molecular modes, occur beyond the light-line in air (red line) $\omega=c |\mathbf{k}|$. The momentum mismatch between the SP and incoming light means that while strong coupling may occur in this system (as an excited SP is not necessary for strong coupling to take place) it cannot be experimentally observed without using one of a variety of experimental tricks to overcome this momentum mismatch. This will be discussed further in the next section. 
\\

As only p-polarised light may excite a SP mode, and constrained by the boundary conditions at the interface, the electric field of the SP for $z>0$ is 
\begin{align}
    \mathbf{E} &=i \left(\frac{\hbar\omega_\mathrm{SP}(k_x)}{2 \epsilon_0 \epsilon_2 V}\right)^{\frac{1}{2}}\\
    &\times\left[ \begin{pmatrix}1\\0\\ i \sqrt{\frac{\epsilon_1(\omega_\mathrm{SP})}{\epsilon_2}}\end{pmatrix}c e^{i k_x x}+ \begin{pmatrix}1\\0\\ -i \sqrt{\frac{\epsilon_1(\omega_\mathrm{SP})}{\epsilon_2}}\end{pmatrix}c^\dagger e^{-i k_x x}  \right]e^{- z/2L_z},
\end{align}
where we have used the fact that $k_z$ is purely imaginary and $ \sim~1 /2 L_z$, where $L_z$ is the penetration depth of the electric field into the dielectric. We consider a sample with dimensions smaller than the propagation length of the SPs in order to neglect the decaying amplitude of the SP in the direction of propagation. The Hamiltonian matrix is given by
\begin{align}
    \mathbf{H} = \begin{pmatrix}
    \omega_\mathrm{SP}(k_x) & g_1(k_x) & g_2(k_x) & \dots & g_N(k_x) \\
    g_1^*(k_x) & \omega_{\mathrm{mol}} & 0 & \dots & 0 \\
    g_2^*(k_x) & 0 & \omega_{\mathrm{mol}}  & \dots & 0 \\
    \vdots & \vdots & \vdots & \ddots & \vdots \\
    g_N^*(k_x) & 0 & 0 & \dots & \omega_{\mathrm{mol}} 
    \end{pmatrix},
\end{align}
%
\begin{figure*}
    \centering
    \includegraphics[width=\textwidth]{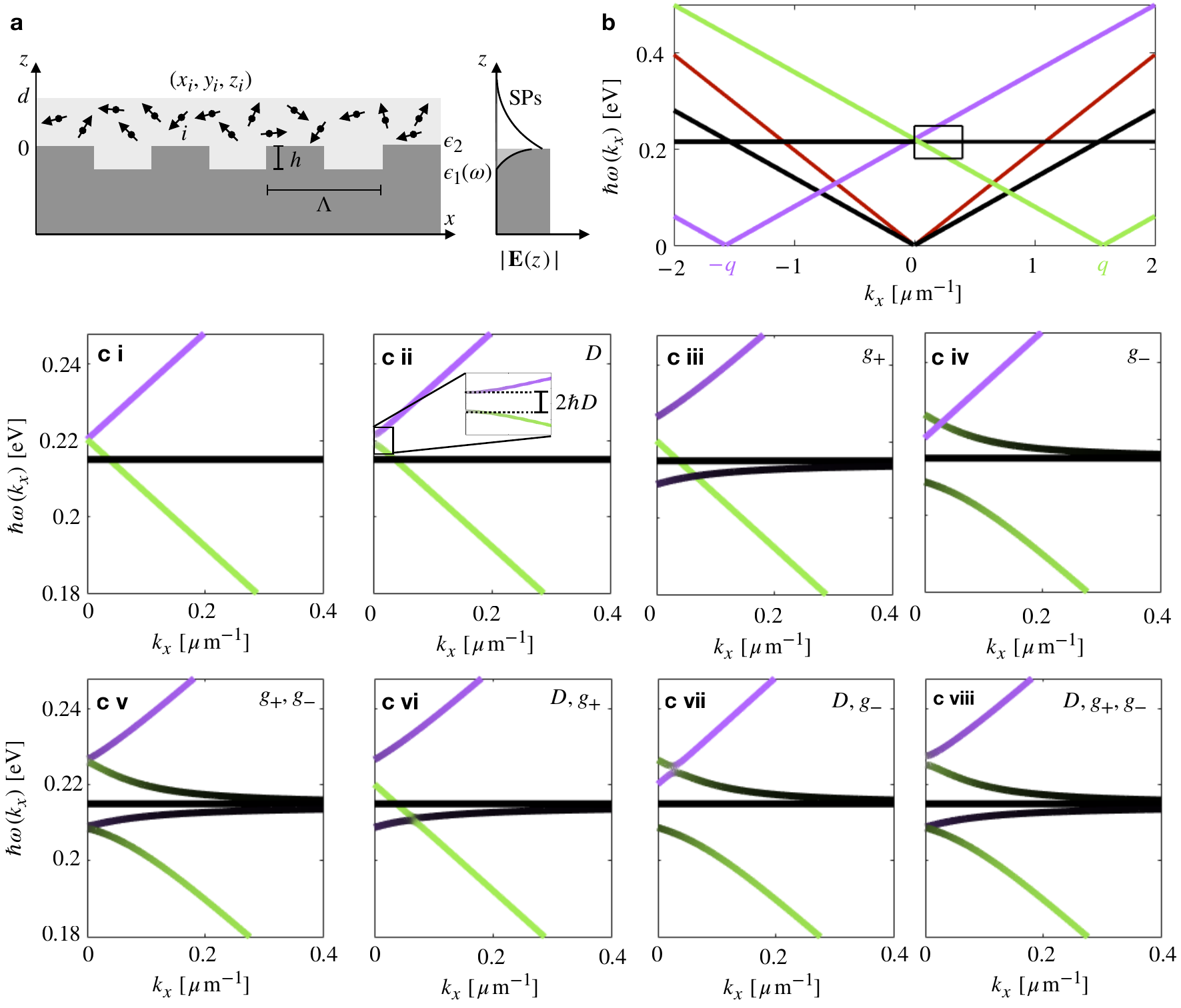}
    \caption{\textbf{Strong coupling of molecules to a surface plasmon on a grating:} \textbf{(a)} Schematic showing a slab of thickness $d$ containing $N$ randomly oriented and distributed dipoles, on a metal grating on which a SP can propagate. \textbf{(b)} The dispersion relation of the uncoupled $0,\pm 1$ plasmon modes on a grating of period $\Lambda=4.0~\mu$m, and $N$ degenerate molecular resonances. The air light-line is given in red. We take $d=0.9~\mu$m. \textbf{(c i)} Zoom-in of the relevant part of the uncoupled dispersion relation, and \textbf{(c ii-viii)} the contribution of each coupling individually and in combination, ending with the complete five-band dispersion relation in \textbf{(c viii)}.}
    \label{fig:SC_molecules_SPP_grating}
\end{figure*}

where the coupling constants are now of the form 
\begin{widetext}
\begin{align}
    g_{i}(k_x) &= i \left(\frac{\omega_{\mathrm{SP}}(k_x)}{2\hbar \epsilon_0 \epsilon_2 V}\right)^{\frac{1}{2}} e^{i k_x x_i- z_i/2L_z}|\mathbf{M}| \left[ \mathrm{cos}\varphi_i \mathrm{sin}\vartheta_i+ i \sqrt{\frac{\epsilon_1(\omega_\mathrm{SP})}{\epsilon_2}}\mathrm{cos}\vartheta_i \right].
\end{align}
\end{widetext}
Diagonalising this Hamiltonian, we find the dispersion relation. As for the planar cavity above, we assume a molecular orientation aligned with the maximal direction of the electric field. The uncoupled system is given in FIG.~\ref{fig:SC_molecules_SPP}b. The section of the plot of interest is expanded in FIG.~\ref{fig:SC_molecules_SPP}c. When coupling is added, we have FIG.~\ref{fig:SC_molecules_SPP}d. As for the case of the cavity, we begin with one photonic mode and $N$ molecular modes, and the result is two hybrid modes and $N-1$ unperturbed molecular modes. As in the previous case, the spatial distribution of the molecules in the $z$ direction causes a reduction in the collective coupling strength. Spatial distribution in the $x$ direction has no effect on the dispersion relation.
\\

As the strong coupling described in this section occurs beyond the light line, it cannot be experimentally probed without the introduction of a tool such as a prism or grating. The addition of a grating is expected to alter the molecule-light interaction. This is what we explore in the next section, which encompasses the main results of this work.
%

\section{Strong coupling of molecules to a surface plasmon on a grating}
\label{sec:SC_molecules_SPP_grating}

The introduction of a grating to the system perturbs the plasmon dispersion relation inside the light-line. This adds some subtleties to the theoretical model, which we now discuss. 
\\

A periodic structure at the dielectric/metal interface on the order of the photonic wavelength (such as the grating with periodicity in the $x$ direction illustrated in FIG.~\ref{fig:SC_molecules_SPP_grating}a) allows for the coherent superposition of transmitted and reflected light in the structure. Through additive interference, the modes which survive the grating are those which have a in-plane momentum corresponding to $k \rightarrow k \pm a 2\pi/\Lambda $, where $\Lambda$ is the period of the grating, and $a=0,1,2,...$ The SP dispersion now presents multiple branches, parts of which will now have been shifted inside the light line. In FIG.~\ref{fig:SC_molecules_SPP_grating}b we illustrate the zero-order $(0)$ and first-order $(\pm1)$ grating-scattered branches of an SP in the presence of a grating with period $\Lambda = 4.0~\mu$m. The close proximity of these SP branches to the molecular resonance $\omega_\mathrm{mol}$ (see FIG.~\ref{fig:SC_molecules_SPP_grating}c i for the relevant zoomed-in section of the dispersion plot) now occurs inside the light-line. The dispersion relation of each new branch is given by, 
\begin{align}
    k_x = \pm q + \frac{\omega_{\mathrm{SP}}}{c}\sqrt{\frac{\epsilon_1 (\omega_{\mathrm{SP}})\epsilon_2 }{\epsilon_1 (\omega_{\mathrm{SP}})+\epsilon_2 }}, 
\end{align}
where we have defined $q=2\pi/\Lambda$. We label the two solutions of $\omega_\mathrm{SP}$ corresponding to $a=\pm 1$ as $\omega_{\mp}$. This labelling convention is chosen so that $\omega_+ (k_x)> \omega_- (k_x)$ for $k_x>0$.  
Coherent superpositions of the $(\pm1)$ scattered SPs may take different spatial positions with relation to the peaks and troughs of the grating~\cite{Barnes_PRB_1996_54_6227}, resulting in a photonic band gap opening at the resonance frequency, such that the plasmon Hamiltonian now reads,
\begin{align}
    \mathcal{H}_{\mathrm{plas}} 
    &= \hbar\omega_+ c_+^\dagger c_+ +  \hbar\omega_- c_-^\dagger c_- +  \hbar D\left[ c_+^\dagger c_- +  c_-^\dagger c_+\right]. 
\end{align}
Each superposition of plasmon modes moves an energy $\hbar D$ away from the unperturbed frequency, where the analytical expression for $D$ is given by~\cite{Barnes_PRB_1996_54_6227},
\begin{align}
\begin{split}
  D &= \frac{c}{2}\Bigg[\sqrt{k_0^2 \left(1-2 s^2\right)+2q^2  s \left(1-3 s^2\right)\bar{\epsilon}^{-1} } \\
  &-\sqrt{k_0^2 \left(1-2 s^2\right)-2q^2  s \left(1-3 s^2\right)\bar{\epsilon}^{-1}}\Bigg].
\label{eq:energy_D}
\end{split}
\end{align}
where $s=qh$, $\bar{\epsilon}=\sqrt{-\epsilon_1 \epsilon_2}$ and $k_0=\omega_0/c$. This results in a total energy gap of $2\hbar D$. For a grating with period $\Lambda=4.0~\mu$m and $h=0.1 \mu$m, equation \ref{eq:energy_D} gives $\hbar D\approx 1.2$~meV. This small splitting can be seen in FIG.~\ref{fig:SC_molecules_SPP_grating}c~ii, in which plasmon-plasmon interactions are considered but the molecules are, for now, considered uncoupled. 

The presence of the grating enforces a periodic boundary condition on the electric field of the SP, such that the plane wave expansions of the first-order ($a=\pm1$) grating-scattered branches are of the modified form $e^{i k_x x} \rightarrow e^{i (k_x\pm q) x}$, such that (for $z> 0$)  
\begin{widetext}
\begin{align}
    \mathbf{E}_{\pm} &=i\left(\frac{\hbar\omega_{\pm}}{2 \epsilon_0 \epsilon_2 V}\right)^{\frac{1}{2}} \left[\begin{pmatrix}1\\0\\ i\sqrt{\epsilon_1(\omega_{\pm})/\epsilon_2}\end{pmatrix}  c_{\pm} e^{i \left(k_x{\mp}q\right)x}- \begin{pmatrix}1\\0\\ -i \sqrt{\epsilon_1(\omega_{\pm})/\epsilon_2}\end{pmatrix}  c_{\pm}^\dagger e^{-i\left( (k_x{\mp}q) x\right)}  \right]e^{- z/2L_z},
\end{align}
\end{widetext}
where $c^{\dagger}_\pm,c_\pm$ are the bosonic creation and annihilation operators for the $\omega_{+}$ and $\omega_{-}$ SPs respectively. The interaction Hamiltonian of the coupled system is given by, 
\begin{align}
    \mathcal{H}_{\mathrm{int}} &= - \sum_{i=1}^N\bm{\mu}_i\cdot \left[\mathbf{E}_++\mathbf{E}_-\right] 
    \\
    \begin{split}
    &= \hbar \sum_{i=1}^N   \bigg[ g^+_{i}(k_x) b^\dagger_i c_+  + \left(g^+_{i}\right)^*(k_x) b_i c_+^\dagger  \\
    &+ g^-_{i}(k_x) b^\dagger_i c_- + \left(g^-_{i}\right)^*(k_x) b_i c_-^\dagger  \bigg] ,
    \end{split}
\end{align}
where the grating-modified interaction strengths are given by,
\begin{widetext}
\begin{align}
    g^{\pm}_{i}(k_x) &= i  \left(\frac{\omega_{\pm}(k_x)}{2\hbar \epsilon_0 \epsilon_2 V}\right)^{\frac{1}{2}}  e^{i(k_x \mp q)x_i}|\mathbf{M}|    \left[ \mathrm{cos}\varphi_i \mathrm{sin}\vartheta_i+ i \sqrt{\frac{\epsilon_1}{\epsilon_2}}\mathrm{cos}\vartheta_i \right]e^{- z_i /2 L_z}.
\end{align}
\end{widetext}

The Hamiltonian can be written in $(2+N)\times(2+N)$ matrix form, 
\begin{align}
    \mathbf{H} = \begin{pmatrix}
    \omega_+(k_x) & D & g^+_1(k_x)& g^+_2(k_x) & \dots & g^+_N(k_x) \\
    D^* &  \omega_-(k_x)  & g^-_1 (k_x) & g^-_2(k_x) & \dots & g^-_N(k_x) \\
    (g^+_1)^*(k_x) & (g^-_1)^*(k_x) & \omega_{\mathrm{mol}} & 0 & \dots & 0 \\
    (g^+_2)^*(k_x) & (g^-_2)^*(k_x) & 0 & \omega_{\mathrm{mol}}  & \dots & 0 \\
    \vdots & \vdots & \vdots & \vdots & \ddots & \vdots \\
    (g^+_N)^*(k_x) &  (g^-_N)^*(k_x) & 0 & 0 & \dots & \omega_{\mathrm{mol}} 
    \end{pmatrix}.
    \label{eq:grating_hamiltonian}
\end{align}

Repeating the same procedure of the previous sections, we find the dispersion relation by diagonalising the Hamiltonian for an explicit example. We use a slab of PMMA of thickness $d=0.9$~$\mu$m and $N/A=70$ nm$^{-2}$ on a gold grating. We consider a sample with dimensions smaller than the propagation length of the SPs so as to neglect the decaying amplitude of the SP. However, unlike the previous system of a planar metal surface, the grating has now introduced an additional length scale. We assume that the spatial distribution of the molecules in the $x$ direction is random, and equally distributed in positions that sample the full grating period. We will return to this point later. Due to the complexity of multiple interactions affecting the dispersion relation, we describe the effects of interactions in the system by `turning on' each interaction one-by-one, illustrated in FIGs.~\ref{fig:SC_molecules_SPP_grating}c~i-viii. FIG.~\ref{fig:SC_molecules_SPP_grating}c~i shows the magnified section of the dispersion relation where the two plasmons are close to the molecular resonance, unaffected by interactions. In FIG.~\ref{fig:SC_molecules_SPP_grating}c ii we include the SP-SP coupling parameterised by $D\neq 0$, in which we see a band gap opening at the crossing point, and a slight mixing of the two modes. In FIG.~\ref{fig:SC_molecules_SPP_grating}c iii we turn on the coupling of the molecules to the upper SP mode, $g_+\neq 0$. We can see the lifting of degeneracy of the molecular modes as one band is pushed away from the upper polariton mode, below the molecular resonance,  with slight mixing of characteristics of the two modes. In FIG.~\ref{fig:SC_molecules_SPP_grating}c iv we see that when $g_-\neq 0$, the degeneracy of the molecular modes is again lifted and a new band is pushed away from the lower polariton, above the molecular resonance. The mixing in this situation is more significant than for when $g_+\neq 0$, due to the direct crossing of the lower polariton with the molecular resonance. FIG.~\ref{fig:SC_molecules_SPP_grating}c v-vii show intermediate systems when two of the couplings are turned on together. When the molecules are coupled to only one polariton branch we see four bands, but when both polaritons are coupled ($g_+,g_- \neq 0$) we see five distinguishable bands, as in FIG.~\ref{fig:SC_molecules_SPP_grating}c v. These are comprised of four bright bands and $N-2$ dark molecular bands. FIG.~\ref{fig:SC_molecules_SPP_grating}c viii shows the final dispersion relation for $D,g_+,g_- \neq 0$, which is very similar to FIG.~\ref{fig:SC_molecules_SPP_grating}c v, with the inclusion of the slightly splitting and mixing of the two plasmon bands at their point of intersection. This model agrees with experimental results~\cite{menghrajani2019vibrational}. 
%
\\
\begin{figure}
    \centering
    \includegraphics[width=\columnwidth]{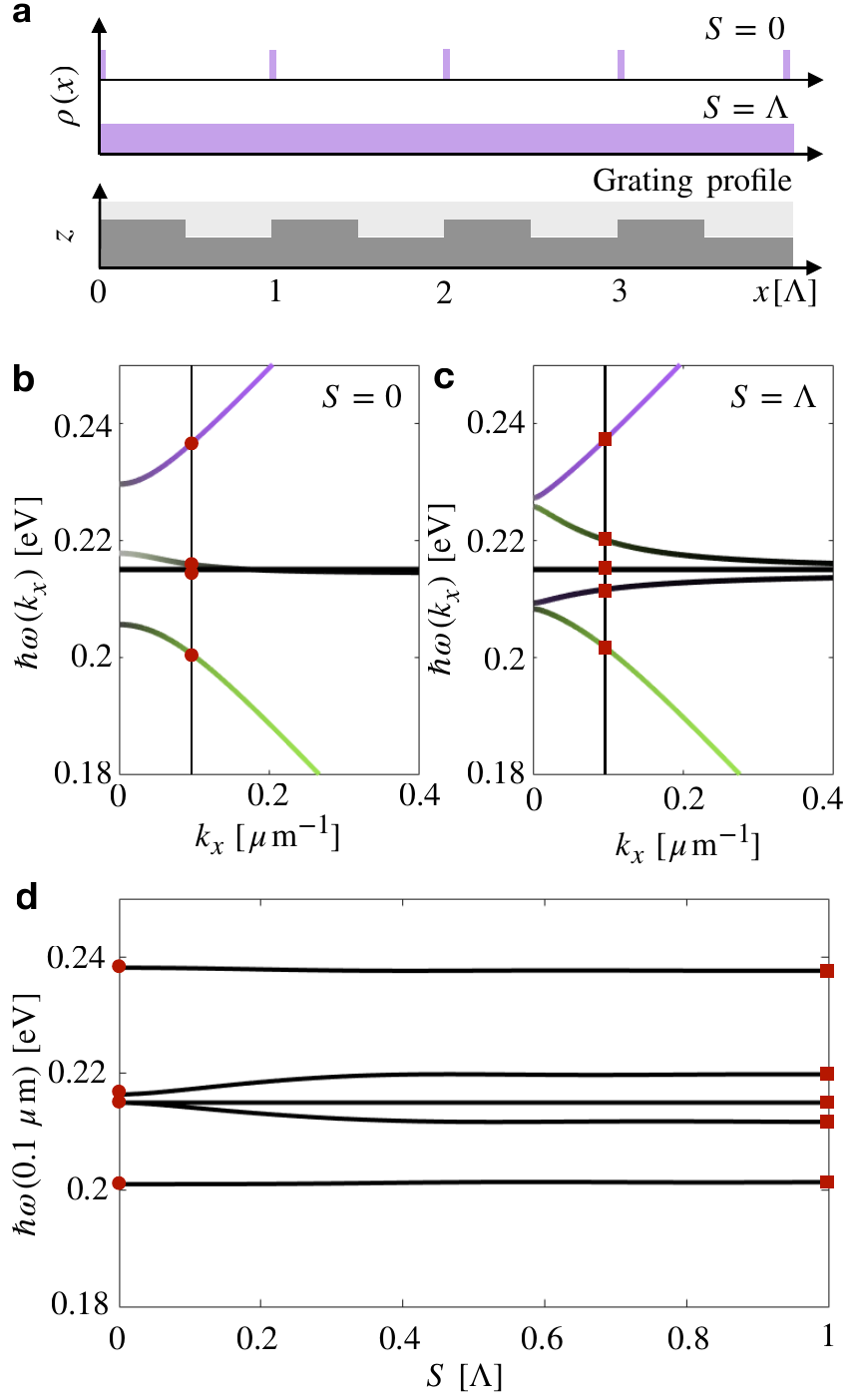}
    \caption{\textbf{Dependency of bands on molecular spatial distribution:} \textbf{(a)} Molecular distribution with $S=0,\Lambda)$ compared to a grating with period $\Lambda$. \textbf{(b)} Dispersion relation with $S=0$, displaying four bands, and the crossing of a polariton mode and the dark molecular modes. \textbf{(c)} Dispersion relation with $S=\Lambda$, displaying five bands and no band crossings. \textbf{(d)} Eigenvalues at $k_x=0.1~\mu$m for varying $S$. For $S>0$ the crossing point of the middle polariton mode and dark molecular resonances becomes an avoided crossing, resulting in a further splitting and an additional mixed (but predominantly molecule-like) mode, and so the four bands split into five bands.}
    \label{fig:variation_spatial}
\end{figure}
\begin{figure*}
    \centering
    \includegraphics[width=\textwidth]{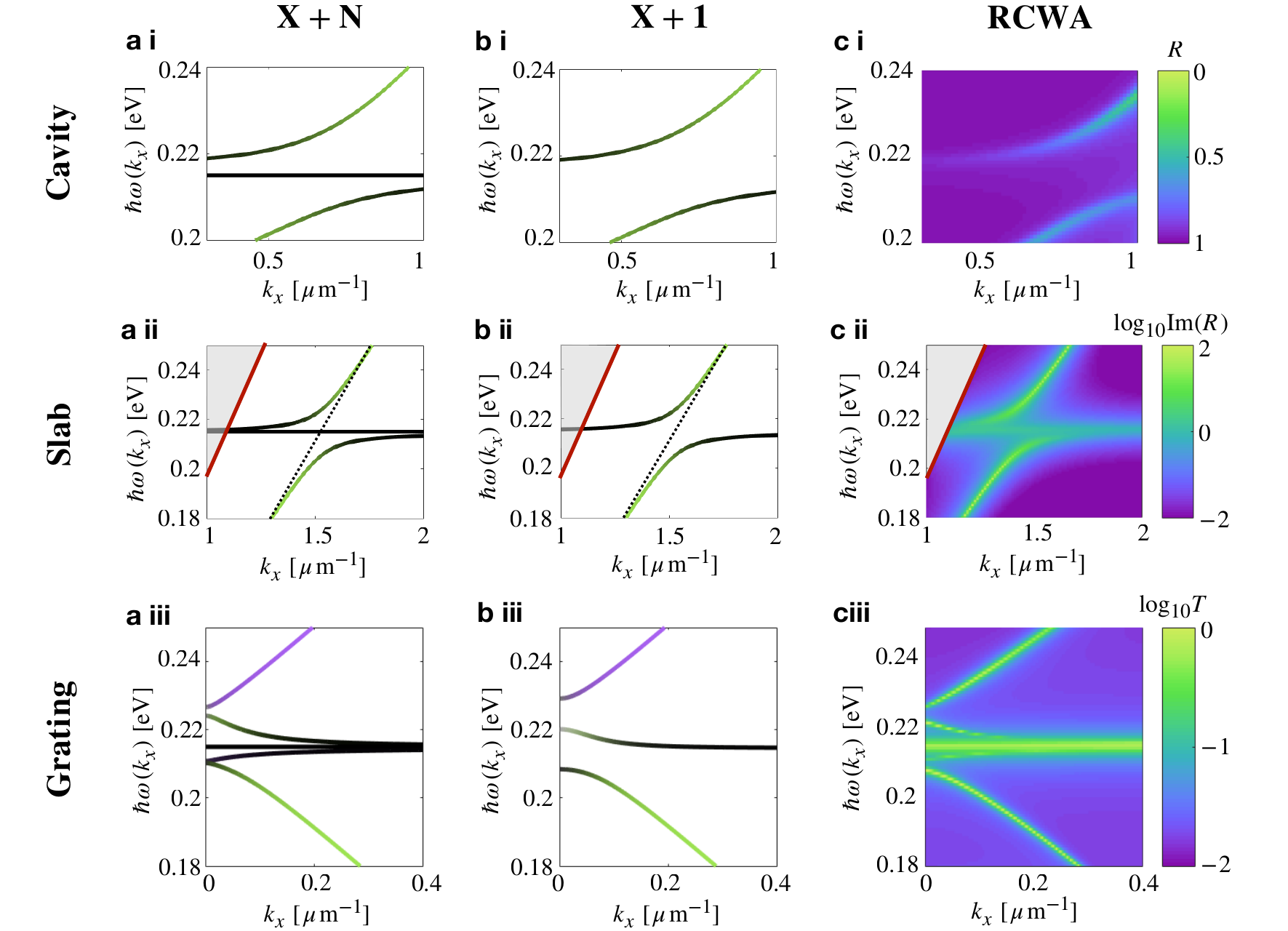}
    \caption{\textbf{Comparing $\mathbf{X+N}$ vs $\mathbf{X+1}$ model:} \textbf{(a)} Dispersion relation for $N$ molecules strongly coupled to \textbf{(i)} the mode of a planar optical cavity, \textbf{(ii)} a SP at a gold/air interface and~\textbf{(iii)} backwards and forwards travelling SPs on a grating. All dispersion relations calculated using a $X+N$ mode. \textbf{(b)} The same dispersion relations calculated using the equivalent $X+1$ models. \textbf{(c)} The same systems simulated using RCWA. The grating in \textbf{c iii} has a period of $4.7~\mu$m as outlined in the text.}
    \label{fig:collective}
\end{figure*}

To see the five-band dispersion relation the molecular distribution and SPs must extend across the full grating profile. The spatial position of each molecule can be defined as $x_i = a_i \Lambda + b_i S$, where $a_i=0,1,2..$, $0\leq b_i \leq 1$, and $0\leq S\leq \Lambda$. $S$ parameterises how molecules are localised in each grating period. For $S=\Lambda$ the molecules are found everywhere, while for $S\rightarrow 0$ the molecules are concentrated in a small fraction of each grating period (FIG~\ref{fig:variation_spatial}a). In FIG.~\ref{fig:variation_spatial} we show that if the molecules occupy a fraction of the grating period $S$ much smaller than one period of the grating, two of the bands in the dispersion relation become degenerate and we observe only four distinct bands as seen in FIG.\ref{fig:variation_spatial}b for $S=0$. When the spatial distribution of molecules extends through the full grating profile, the breaking of finite translational invariance $x_i\rightarrow x_i + \Lambda$ results in the onset of a fifth distinct band in the dispersion relation as seen in FIG.~\ref{fig:variation_spatial}c for $S=\Lambda$. In FIG.~\ref{fig:variation_spatial}d we plot the eigenvalues of the system Hamiltonian at $k_x=0.1~\mu$m for $0\leq S\leq \Lambda$. We can see that as $S\rightarrow 0$, the $N-4$ dark resonances and the lower polariton closest to the molecular resonance reduce to $N-3$ dispersionless molecular modes. The molecules must extend across approximately half of the grating profile ($S\approx 0.4 \Lambda$) to recover the approximate eigenvalues for the system of homogeneously distributed molecules on a grating. The variation in band positions as $S$ is varied suggests that there are optimal ways in which one can choose the spatial distribution of the molecules with respect to the grating profile, in order to tune the band separation\cite{Barnes_PRB_1996_54_6227}. 
\\

The final point of this paper is to compare the results of a $X+N$ model in which $X$ photonic modes and $N$ molecular resonances are individually included in the Hamiltonian, vs the results of a $X+1$ model in which the molecules are treated via their collective behaviour as the collective coupling strength $g_N = \sqrt{\sum_i{|g_i|^2}}\propto \sqrt{N/A}$. This model usually captures the bright modes of the system, allowing one to calculate important properties of the system such as the Rabi frequency. We compare both models to rigorous coupled wave analysis (RCWA ~\cite{manceau_zanotto2013}). For the RCWA, we model PMMA using a bulk dielectric function
\begin{align}
    \epsilon_\mathrm{PMMA} &= \epsilon_2 + \frac{A \omega_0^2}{\omega^2-\omega_0^2+i \gamma \omega },
\end{align}
where $\epsilon_2=1.99$, $\omega_0 = 3.28\times 10^{14}~$rad s$^{-1}$, $\gamma = 2,45\times 10^{12}~$rad s$^{-1}$ and $A = 0.0165$~\cite{shalabney2015coherent}. The simulation was performed with a half number of harmonic waves of 60.
In FIG.~\ref{fig:collective}a we repeat the calculated dispersion relations for the three systems discussed in this work - (a i) molecules strongly coupled to the mode of a planar cavity, (a ii) molecules strongly coupled to a plasmon at a dielectric/metal interface and (a iii) molecules strongly coupled to forward and backwards scattered plasmons on a grating, using the same system parameters as the previous sections. In FIG.~\ref{fig:collective}b~i-iii we model the same systems but the $N$ molecular modes are treated here via their collective mode. In the case of  the cavity (b i), the qualitative behaviour of the polariton modes is the same as for the $X+N$ model, but the dark modes are absent. In the case of the plasmon on a slab (b ii), the $1+1$ model gives the correct quantitative description of the bright states, but again the dark states are not captured. In the case of the grating system (b iii), using a $2+1$ model, three of the bright bands are captured well although with slightly different band gap values. The dark molecular resonance is not seen, and neither is the nearly-molecule-like bright band below the dark molecular resonance. The full basis of states - using all states in the single particle basis as described in this work, or using a basis of collective states including both bright and dark modes - is required to capture the behaviour of molecules interacting with plasmonic modes on a grating, and will likewise also be necessary for other systems with additional features such as defects or other properties which break translational symmetry of the system. FIG~\ref{fig:collective}c~i-iii gives the results of RCWA simulations of the three systems. For (c i) the cavity, we model the system as two gold mirrors of thickness $0.01~\mu$m, fully filled with a slab of PMMA of thickness $2.26~\mu$m. We see that reflectivity calculated using RCWA reproduces the bright modes of the cavity. The molecular dark states are not seen in the simulation due to high impedance mismatch that prevents much light entering the cavity around the resonance frequency. The $X+1$ model is thus a good description of what is seen in cavity experiments. For (c ii) the slab system, we model the $2.0~\mu$m slab of PMMA on top of a $0.1~\mu$m slab of gold, sitting on a substrate of CaF$_2$ of thickness 30~$\mu$m. We have plotted the imaginary component of the reflection in order to observe physics below the light line. The dark states  can clearly be seen in the dispersion relation, and so the $X+N$ model gives a more complete description of the system. In ~\ref{fig:collective}c iii we plot the RCWA simulated dispersion relation for the grating system. We have modelled the gold grating as a slab of thickness $0.2~\mu$m, and matter is removed to a depth of $0.1~\mu$m with periodicity $4.7~\mu$m, and slots of width $1~\mu$m. The gold sits in a substrate of CaF$_2$ of depth $30~\mu$m, and the PMMA on top of the grating has thickness $0.9~\mu$m. The $X+N$ model gives good agreement, although the analytical result found for grating period $\Lambda=4.0~\mu$m best fits to the calculated system with grating period of $\Lambda = 4.7~\mu$m, owing to the simplified description of the grating used in the analytical model.  

To see clearly what is happening in this system we have assumed SPs exist only on one side of the grating. However, for a sufficiently thin metal film we would expect coupled SPs on both the top and bottom of the grating, resulting in a more complex dispersion diagram and optical spectra of the  system. This, as well as additional interesting and important features such as non-zero line-widths, molecular interactions and inhomogeneous broadening could all be added to the simple model we have presented here. The grating in this work has been treated as a simple periodic boundary condition on the plasmon electric field. While this approach gives good agreement with numerical simulations, the specific form of the grating profile will have an affect on plasmon-molecule coupling and should eventually be taken into account. The tuning of the electric near-field profile will provide an additional tool for controlling plasmon-molecule interactions. The work presented in this paper uses a $(X+N)\times(X+N)$ representation of the Hamiltonian matrix. For large values of the molecular dipole moment (equivalently the oscillator strength in a macroscopic model) or particularly large mode volumes, the system moves into the intermediate strong coupling regime, and a $(2XN)\times(2XN)$ model\footnote{The two models are usually referred to as $N+1$ vs $2N$, where $N$ is the number of photonic modes, coupled to a single molecular mode. The current work uses different notation, i.e. $X$ photonic modes and $N$ molecular modes.} is required to accurately reproduce the dispersion relation~\cite{richter2015maxwell,balasubrahmaniyam2021coupling}. This would be a straightforward extension of the current work using the same coupling coefficients described in this paper. 

\section{Conclusions}
\label{sec:conclusions}

In this work we have developed the theoretical model of molecules interacting with surface plasmons in the presence of a grating. The main result of this work is the $(2+N)\times(2+N)$ Hamiltonian~\ref{eq:grating_hamiltonian} of the system, with spatial- and grating-dependence, which when diagonalised gives a five-band dispersion relation. Of these five distinguishable bands, four bands are bright states and $N-2$ are dark, collective molecular states. For a single electromagnetic mode coupled to molecules we expect to see three distinct bands. For two electromagnetic modes coupling to molecules we expect to see four distinct bands, and the breaking of translational invariance with the addition of a grating adds a fifth band. While it is already known that changing the grating period will change the dispersion relation of the system, by calculating the Hamiltonian of the system from a first principles approach using the full $2+N$ system, we can calculate the full $(2+N)$-band dispersion relation and also demonstrate the importance of the spatial distribution of the molecules.  We show that an $X+1$ model does not qualitatively or quantitatively describe the features of this system, as compared to RCWA simulations, as the full basis of states is required. With ever improving techniques to control the orientation and placement of molecules on surface, the spatial distribution of the molecules may be an additional tool in engineering the bands of molecule-light hybrid systems. 

\section*{Acknowledgements}
The authors would like to acknowledge many useful discussions with Kishan Menghrajani. 
JB acknowledges support from European Research Council (ERC) under Horizon 2020 research and innovation programme PICOFORCE (Grant
Agreement No. 883703), THOR (Grant Agreement No. 829067) and POSEIDON (Grant Agreement No. 861950). JB acknowledges funding from the EPSRC (Cambridge NanoDTC EP/L015978/1, EP/L027151/1, EP/S022953/1, EP/P029426/1, and EP/R020965/1). WLB and MSR acknowledge funding from the Molecular 
Photonic Breadboards grant EP/T012455/1. WLB acknowledges funding from the Leverhulme Trust and ERC through the photmat project (ERC-2016-AdG-742222, www.photmat.eu). R.A. acknowledges support from the Rutherford Foundation of the Royal Society of New Zealand Te Apārangi, the Winton Programme for the Physics of Sustainability, and Trinity College Cambridge.
\newpage
\bibliographystyle{unsrt}
\bibliography{references}
\newpage
\appendix

\section{Low-excitation approximation}
\label{app:low_ex}

The excitation of a molecule (either excitonic or vibronic) can be described as a two level system with energy levels $|g\rangle$ and $|e\rangle$, given in vector form as 
\begin{align}
    |g\rangle = \begin{pmatrix}0\\1\end{pmatrix},\quad |e\rangle = \begin{pmatrix}1\\0\end{pmatrix}.
\end{align}
The transition energy is given by $\hbar \omega_{\mathrm{mol}} = E_e-E_g$. The transition dipole moment can be expanded in terms of raising and lowering operators such that the dipole operator of the $i$-th molecule is given by 
\begin{align}
  \bm{\mu}_i &= \mathbf{M}_i\bm{\sigma}^+_i + \mathbf{M}^*_i\bm{\sigma}^-_i
\end{align}
where $\mathbf{M}_i=e_0 \langle e| \mathbf{d}_i|g\rangle$ is the $i$-th dipole matrix element, $\mathbf{d}_i$ is the displacement vector of the $i$-th dipole, and the raising and lowering operators
\begin{align}
    \bm{\sigma}_i^+ = \begin{pmatrix}0&1\\0&0\end{pmatrix},\quad \bm{\sigma}_i^- = \begin{pmatrix}0&0\\1&0\end{pmatrix}
\end{align}
act on the $i$-th molecule. By performing a Holstein-Primakoff transformation we may replace the fermionic raising and lowering operators, $\sigma_i^+$ and $\sigma_i^-$, with bosonic operators, $b_i^\dagger$ and $b_i$, which act on the Fock state for the excitation of the $i$-th molecular as $|n_i\rangle$. We restrict ourselves to the low-excitation regime, such that $n_i=0,1$. The molecular dipole is now written as 
\begin{align}
    \bm{\mu}_i &= \mathbf{M}_i b^\dagger_i+ \mathbf{M}^*_i b_i.
\end{align}

\end{document}